\documentstyle[12pt]{article}

\begin{document} 

\begin{titlepage}

\hrule 
%%%%%%%%%%%%%%%%%%%%
\leftline{}
\leftline{Preprint 
          \hfill   \hbox{\bf CHIBA-EP-94}}
\leftline{\hfill   \hbox{\bf OUTP-96-30P}}
\leftline{\hfill   \hbox{hep-th/9608100}}
\leftline{\hfill   \hbox{June 1996}}
\vskip 5pt
%\leftline{}
\hrule 
%%%%%%%%%%%%%%%%%%%%
\vskip 1.5cm

\centerline{\large\bf 
Transverse Ward-Takahashi Identity, Anomaly
}
\centerline{\large\bf
and Schwinger-Dyson Equation
$^*$
}   
\vskip 1cm

\centerline{{\bf 
Kei-Ichi Kondo$^{\dagger}$
}}  
\vskip 4mm
\begin{description}
\item[]{\it  %$^1$ 
  Theoretical Physics, 
  %Department of Physics, 
  University of Oxford,
  1 Keble Road, Oxford, OX1 3NP, UK.$^\ddagger$
  }
\item[]{$^\dagger$ 
  E-mail: kondo@cuphd.nd.chiba-u.ac.jp;
  kondo@thphys.ox.ac.uk}    
\end{description}
\vskip 1cm

\centerline{{\bf Abstract}} \vskip .5cm

Based on the path integral formalism, we rederive and
extend the transverse Ward-Takahashi identities (which were
first derived by Yasushi Takahashi) for the vector and the
axial vector currents and simultaneously discuss the
possible anomaly for them.   Subsequently, we propose a new
scheme for writing down and solving the Schwinger-Dyson
equation in which the the transverse Ward-Takahashi
identity together with the usual (longitudinal)
Ward-Takahashi identity are applied to  specify the
fermion-boson vertex function.   Especially, in two
dimensional Abelian gauge theory, we show that this scheme
leads to the exact and closed Schwinger-Dyson equation for
the fermion propagator in the chiral limit (when the bare
fermion mass is zero) and that the Schwinger-Dyson equation
can be exactly solved.


PACS numbers:

\vskip 0.5cm
\hrule  
%%%%%%%%%%%%%%%%%%%%%%%%%%%%%%
%\vskip 2cm  
%\hrule  
%\bigskip  
%\centerline
%{\bf CHIBA UNIVERSITY}  
%\vfill 

%\vskip 2cm  
%\hrule  

\begin{description}
\item[]{
$^\ddagger$
Address from March 1996 to December 1996.
  On leave of absence from: 
  Department of Physics, Faculty of Science,
  Chiba University, Chiba 263, Japan.
  }
%\item[]{
%$^*$ 
% To be published in .
% Submitted to .
% }  
\end{description}

\end{titlepage}

%\newpage
%%%%% Table of Contents %%%%%
\pagenumbering{roman}
%\tableofcontents
%%%%% Table of Contents %%%%%
\pagenumbering{arabic}
%%%%%

\newpage

\section{Introduction} 
\setcounter{equation}{0}

If a quantum field theory possesses some symmetry, there
exist various identities among Green functions of the
theory.  They are in general called the Ward-Takahashi (WT)
identities
\cite{WT}. If the action is invariant under the continuous
symmetry (i.e. transformation with continuous parameters)
of the dynamical variable, this inevitably leads to the
existence of the corresponding conservation law, as  
Noether's theorem claims. However, the symmetry in the
classical theory (at tree level) may be broken in the
quantum theory (at loop level), which is the phenomenon
known as the anomaly
\cite{ABJ69}. If we could construct all the correlation
functions (more precisely, the Wightman function) satisfying
all the (infinite number of) WT identities associated with
some symmetry, we would be in principle able to recover the
original theory with the symmetry in question.
\par
For recent two decades, this kind of attempt has been
extensively performed in the framework of the
Schwinger-Dyson (SD) equation 
\cite{SD} for the purpose of studying
the strong coupling phase of the gauge theory
\cite{MN74,FK76,Miransky85}.
The strong coupling phase of QED is believed to exist
above the critical coupling $e_c$ of order unity, which
is the phase transition point accompanied by the
spontaneous breakdown of chiral symmetry \cite{Kogut88}.
It is notorious that the simple approximation (called ladder
or rainbow approximation) leads to the severely
gauge-dependent result for the gauge-invariant quantity. 
This is because we must truncate the infinite series of SD
equations so that we are able to handle with them in
actually solving the SD equation and the appropriate
procedure of truncating the series in a gauge-invariant
manner is not known, although, in the perturbation theory,
the gauge-invariance is preserved order by order by using
the gauge-invariant regularization. 
In light of this, the fermion-boson (photon) vertex function
$\Gamma_\mu(x,y;z)$ is the most difficult quantity to be
specified as we first encounter in the framework of
the SD equation of gauge theory.  In recent several years,
however,  there have been considerable efforts to improve
the vertex so that it satisfies the WT identity as a result
of gauge invariance, although there are infinite number of
WT identities coming from the gauge invariance.  The
longitudinal part of
$\Gamma_\mu$  can be written in terms of the full fermion
propagator $S$, that is to say, in momentum space
\begin{eqnarray}
  k_\mu \Gamma^\mu(q,p) = S^{-1}(q) - S^{-1}(p) ,
  \quad
  k_\mu := q_\mu - p_\mu .
 \label{WTL}
\end{eqnarray}
This observation might suggest a possibility of writing down
the self-consistent and closed SD equation for the full
fermion propagator. Recently it has been fully recognized
that the usual WT identity is not sufficient to specify the
vertex function in the non-perturbative study, actually it
does not at all specify the transverse part of the vertex.
Some people have tried to determine or choose the most
plausible form for the vertex by requiring a number of
conditions: no kinematic singularity, multiplicative
renormalizability, agreement with the lower order
perturbation theory, etc. 
\cite{Roberts94,Atkinson94,Pennington95}.
Such a sort of attempt reached to the top quite recently at
least in the quenched case (i.e. the limit of zero
fermion-flavor or no vacuum polarization to the
photon propagator)
\cite{Pennington96}.
The SD equation for the (3-point) vector vertex function is
not closed.  This is one of the reasons why the gauge theory
defined on space-time in dimensions greater than two can
not be solved exactly.  Moreover, the vector vertex
function may have various types of tensor structures
\cite{BC80}, in sharp contrast with the full fermion and
the full photon propagators (2-point functions).  This fact
makes the actual analysis more difficult.
\par
An aim of this paper lies in providing more information
which is useful to specify the transverse vertex based
on a new type of WT identity
\cite{Takahashi86}, which is called the transverse WT
identity.  In contrast, the usual WT identity is 
called the longitudinal WT identity.
The transverse WT identity was derived by Takahashi more
than ten years ago \cite{Takahashi86}.  In this paper, we
rederive it from the path integral formalism and examine the
possibility of the anomaly associated with it.
The transverse WT identity
specify the {\it rotation} of the vertex
\begin{eqnarray}
 \partial_\mu \Gamma_\nu - \partial_\nu \Gamma_\mu ,
 \label{rot}
\end{eqnarray}
while the longitudinal WT identity determines its divergence
\begin{eqnarray}
 \partial^\mu \Gamma_\mu . 
 \label{div}
\end{eqnarray}
We point out that two types of combination (\ref{rot})
and (\ref{div}) are enough to specify the vertex uniquely
in the SD equation for the fermion propagator, although
they do not specify the vertex function itself.   It should
be kept in mind that the transverse WT identity is not
closed in itself, since the rotation can not in general be
written only in terms of the fermion propagator.  In this
point, the transverse part is quite different from the
longitudinal part. Nevertheless, we show that, in two
dimensional Abelian gauge theory, the transverse WT identity
gives the closed set of SD equations (in the limit of zero
bare fermion mass) together with the two propagators, and
a set of SD equation can be exactly soluble. In $D>2$
dimensions, such a simple situation does not occur.  But
the truncated transverse WT identity in the same form as
the two-dimensional case leads to the exactly soluble SD
equation for the fermion propagator in $D>2$ dimensions
without any further approximation (linearization and
separation of the kernel for the angular variable). This
may be an alternative starting point in looking for the
appropriate ansatz for the vertex.
\par
This paper is organized as follows.
As a preliminary to the subsequent sections, in section 2
we review the usual WT identity for fixing the notation.  In
section 3, we rederive the transverse WT identity based on
the path integral formalism.  Our presentation allows
straightforward extension to the non-Abelian gauge theory.
In section 4, the transverse WT identity
for the axial vector current is derived in the same way as
in section 3. In these two sections, 3 and 4, we neglect the
additional term which may come from the anomaly.  The
anomaly is taken into account in section 5 based on
the Fujikawa's method
\cite{Fujikawa79} in which the anomaly is identified with
the Jacobian accompanied by the transformation of fermionic
variables  in the path integral measure. This gives an
alternative derivation of the transverse WT identity as
well as the longitudinal WT identity. In section 6, we
apply the transverse WT identity to the SD equation for the
fermion propagator. We propose a new strategy to specify the
SD equation. In section 7, we show that in 1+1 dimensions
the transverse and the longitudinal WT identities can lead
to the exact and closed SD equation and the SD equation
derived in such a way can be exactly solvable.  In the final
section we summarize the result.

%\newpage
\section{Usual longitudinal WT identity} 
\setcounter{equation}{0}

We consider the theory with the Lagrangian
\begin{eqnarray}
  {\cal L}[\bar \Psi, \Psi, A] 
  &=& {\cal L}_{F}[\bar \Psi, \Psi, A] 
  + {\cal L}_{g}[A] ,
  \nonumber\\ 
{\cal L}_{F}[\bar \Psi, \Psi, A] 
&:=&   \bar \Psi i \gamma^\mu (\partial_\mu - ie A_\mu)\Psi
  - \bar \Psi M \Psi,
\end{eqnarray}
where ${\cal L}_{g}$ is the Lagrangian for the gauge
field specified below,  $\Psi$ is the Dirac fermion
with a spinor index
$\alpha$ and a flavor index $i$ (and possibly a color index
for non-Abelian gauge theory) and $M$ is a mass
matrix for the fermion. 
For a while, we restrict our attention to the Abelian
gauge case and define the vector current
${\cal J}_\mu$ by
\footnote{In the following we do not write the indices
explicitly when the indices are contracted in a obvious way.
In this paper $A:=B$ implies that $A$ is defined by $B$.
}
\begin{eqnarray}
{\cal J}_\mu(x) := \bar \Psi(x) \gamma_\mu \Psi(x) 
=  \bar \Psi_{\alpha}^i(x) (\gamma_\mu)^{\alpha\beta}
\Psi_{\beta}^i(x) .
\end{eqnarray}
It is well known that the usual WT identity is derived from
\begin{eqnarray}
 {\cal D}(x)_J
 := \partial_\mu \langle {\cal J}^\mu(x) \rangle_J
 = i \langle \bar \Psi(x) \eta(x) \rangle_J
 - i \langle \bar \eta(x) \Psi(x) \rangle_J,
 \label{WTL0}
\end{eqnarray}
where we have defined the expectation value 
$\langle ... \rangle_J$ in the presence of a set of
external sources  
$J:=\{ J^\mu, \eta, \bar \eta \}$ by using the functional
integral:
\begin{eqnarray}
  \langle {\cal O}(x) \rangle_J
 &=&  {[[ {\cal O}(x) ]]_J \over [[1]]_J},
\end{eqnarray}
\begin{eqnarray}
[~[ {\cal O}(x) ]~]_J
 &=& \int {\cal D}\bar \Psi {\cal D}\Psi {\cal D}A_\mu
 \exp \left[i \int d^D x({\cal L}+{\cal L}_{J}) 
 \right]
 {\cal O}(x),
\end{eqnarray}
with Schwinger's source term:
\begin{eqnarray}
  {\cal L}_{J} :
  =  A_\mu(x) J^\mu(x) + \bar \Psi(x) \eta(x)
  + \bar \eta(x) \Psi(x)  .
\end{eqnarray}
Here note that the external sources have possible indices
corresponding to those of the field.

\par

On the other hand, 
if we consider the Abelian gauge theory whose Lagrangian
(with a gauge-fixing term) is given by
\begin{eqnarray}
 {\cal L}_{g}[A] 
 =  - {1 \over 4} F_{\mu\nu}F^{\mu\nu}
   - {1 \over 2\xi}(\partial^\mu A_\mu)^2 ,
\end{eqnarray}
the identity
\begin{eqnarray}
  \int {\cal D}\bar \Psi {\cal D}\Psi {\cal D}A_\mu
  {\delta \over \delta A_\mu}
  \exp \left[i \int d^D x({\cal L}+{\cal L}_{J}) 
  \right]
 \equiv 0,
 \label{ID1}
\end{eqnarray}
leads to 
\begin{eqnarray}
  \langle {\cal J}^\mu(x) \rangle_J
 =  - {1 \over e}
 \langle \Delta^{\mu\rho}(\partial) A_\rho(x) + J^\mu
\rangle_J,
 \label{WTA0}
\end{eqnarray}
where $\Delta_{\mu\rho}(\partial)$ is the inverse of the
free boson propagator
$D_{\mu\rho}^{(0)}(\partial)$:
\begin{eqnarray}
 \Delta_{\mu\rho}(\partial) 
 := D_{\mu\rho}^{(0)}{}^{-1}(\partial)
 = \partial^2 g_{\mu\rho} - \partial_\mu \partial_\rho
 + \xi^{-1} \partial_\mu \partial_\rho .
\end{eqnarray}
\par
Now we introduce the generating functional for the
connected correlation function: 
\begin{eqnarray}
 i W(J^\mu, \eta, \bar \eta) 
 := \ln Z(J^\mu, \eta, \bar \eta) ,
\quad
 Z(J^\mu, \eta, \bar \eta) = [[ 1 ]]_J .
\end{eqnarray}
Then Eq.~(\ref{WTA0}) reads
\begin{eqnarray}
  \langle {\cal J}^\mu(x) \rangle_J
 =  - {1 \over e} \left[ \Delta^{\mu\rho}(\partial) 
 {\delta W \over \delta J^\rho} + J^\mu \right],
 \label{WTA1}
\end{eqnarray}
since
\begin{eqnarray}
  {\delta W \over \delta J^\mu(x)} 
  =  \langle A_\mu(x) \rangle_J,
  \quad
  {\delta W \over \delta \eta(x)} 
  = - \langle \bar \Psi(x) \rangle_J,
  \quad
  {\delta W \over \delta \bar \eta(x)} 
  =  \langle \Psi(x) \rangle_J .  
 \label{g}
\end{eqnarray}
From Eq.~(\ref{WTA1}), we obtain
\begin{eqnarray}
 \partial_\mu \langle {\cal J}^\mu(x) \rangle_J
&=& - {1 \over e} \left[ \partial_\mu
\Delta^{\mu\rho} (\partial) 
{\delta W \over \delta J^\rho} + \partial_\mu J^\mu(x)
 \right]
 \nonumber\\
&=& - {1 \over e}
 \langle \partial_\mu \Delta^{\mu\rho}(\partial) A_\rho(x) 
 +  \partial_\mu J^\mu  \rangle_J,
 \label{WTL3}
\end{eqnarray}
where only the longitudinal part of the (inverse) boson
propagator contributes to the right-hand-side of this
equation. Note that Eq.~(\ref{WTL0}) is rewritten as
\begin{eqnarray}
 \partial_\mu \langle {\cal J}^\mu(x) \rangle_J
 = - i {\delta W \over \delta \eta(x)} \eta(x)  
 - i \bar \eta(x) {\delta W \over \delta \bar \eta(x)} .
 \label{WTL1}
\end{eqnarray}
Then, differentiating Eq.~(\ref{WTL3}) and Eq.~(\ref{WTL1})
with respect to $\bar \eta(y)$ and $\eta(z)$ and
then putting 
$J^\mu = \eta = \bar \eta = 0$, we obtain
\begin{eqnarray}
 {\cal D}(x,y,z) &:=&
\partial_\mu \langle {\cal J}^\mu(x) ;
 \Psi(y) ; \bar \Psi(z) \rangle_c
  \nonumber\\
&=&  \langle {1 \over e} 
\partial_\mu \Delta^{\mu\rho} (\partial) A_\rho(x) ;
 \Psi(y) ; \bar \Psi(z) \rangle_c
  \nonumber\\
 &=&  \langle \Psi(y) \bar \Psi(z) \rangle_c \delta^D(x-z)
 -  \langle \Psi(y) \bar \Psi(z) \rangle_c \delta^D(x-y),
\end{eqnarray}
where $\langle ...\rangle_c$ denotes the connected
correlation function.
For the proper fermion-boson vertex function in momentum
representation

\begin{eqnarray}
 S(q)\Gamma^\mu(q,p)S(p) := \int d^Dy \int d^Dz 
 e^{i(q \cdot y-p \cdot z)} 
 \langle {\cal J}^\mu(0) ;
 \Psi(y) ; \bar \Psi(z) \rangle_c,
 \label{FT1}
\end{eqnarray}
the well-known form of the WT identity is recovered:
\begin{eqnarray}
  k_\mu \Gamma^\mu(q,p) = S^{-1}(q) - S^{-1}(p) ,
  \quad
  k_\mu := q_\mu - p_\mu .
 \label{WTL5}
\end{eqnarray}
\par
The full gauge-boson propagator 
$D_{\mu\nu}(\partial)$ obeys the SD equation:
\begin{eqnarray}
D_{\mu\nu}^{-1}(\partial)
= D_{\mu\nu}^{(0)}{}^{-1}(\partial) -
\Pi_{\mu\nu}(\partial) .
\end{eqnarray}
In gauge theory, the longitudinal part of the full
gauge-boson propagator is the same as that of the free one,
since
\begin{eqnarray}
 \Pi_{\mu\nu}(\partial)  = (g_{\mu\nu}\partial^2 - 
 \partial_\mu \partial_\nu) \Pi(\partial),
\end{eqnarray}
due to gauge invariance.
Therefore, we obtain
\begin{eqnarray}
 \partial^\mu D_{\mu\rho}^{(0)}{}^{-1}(\partial)
 = \xi^{-1} \partial^2 \partial_\rho  
 = \partial^\mu D_{\mu\rho}{}^{-1}(\partial) .
\end{eqnarray}
Hence this property can be also derived as a consequence of
the WT identity.
\footnote{
If the vector boson has the mass $\mu$, the
inverse of the free propagator reads
$
 D_{\mu\rho}^{(0)}{}^{-1}(\partial)
 = \partial^2 g_{\mu\rho} - \partial_\mu \partial_\rho
 + \xi^{-1} \partial_\mu \partial_\rho 
 + \mu^2 g_{\mu\rho} .
$
Usually, the naive introduction of such a term breaks the
gauge invariance and hence it seems at first glance that
this property can not be preserved.  However, if we
introduce the additional scalar freedom (St\"uckelberg
field) and adopt the R$_\xi$ gauge, we can extend this
scheme into the massive gauge-boson model. This issue will
be discussed in detail in a subsequent paper
\cite{Kondo96c}. 
}
\par
Furthermore, we introduce the generating functional for
the one-particle irreducible diagram through the Legendre
transform
\begin{eqnarray}
 \Gamma(\bar \Psi, \Psi, A_\mu) 
 := W(\eta, \bar \eta, J^\mu) -
\int d^Dx[A_\mu(x)J^\mu(x) 
 + \bar \Psi(x)\eta(x) + \bar \eta(x) \Psi(x)] .
 \label{Legendre}
\end{eqnarray}
By making use of the relations following from
(\ref{Legendre}):
\begin{eqnarray}
  {\delta \Gamma \over \delta A_\mu(x)} 
  = -  J^\mu(x) ,
  \quad
  {\delta \Gamma \over \delta \bar \Psi(x)} 
  = -  \eta(x) ,
  \quad
  {\delta \Gamma \over \delta \Psi(x)} 
  =   \bar \eta(x) ,  
 \label{1PI}
\end{eqnarray}
Eq.~(\ref{WTA0}) is further rewritten as
\begin{eqnarray}
  \langle {\cal J}^\mu(x) \rangle
 =    - {1 \over e} \left[ \Delta^{\mu\rho}(\partial) 
  A_\rho(x)_J 
 - {\delta \Gamma \over \delta A^\mu(x)} \right],
 \label{WTA2}
\end{eqnarray}
while Eq.~(\ref{WTL0}) is rewritten as
\begin{eqnarray}
 \partial_\mu \langle {\cal J}^\mu(x) \rangle_J
 = - i \bar \Psi(x) {\delta \Gamma \over \delta \bar \Psi(x)}
 - i  {\delta \Gamma \over \delta \Psi(x)} \Psi(x) .
 \label{WTL2}
\end{eqnarray}
Substituting Eq.~(\ref{WTA2}) into Eq.~(\ref{WTL2}), we
obtain
\begin{eqnarray}
  - {\cal D}(x)_J = {1 \over e} \left[ \partial_\mu
\Delta^{\mu\rho} (\partial) 
  A_\rho(x)
 - \partial_\mu {\delta \Gamma \over \delta A^\rho(x)} 
 \right]
 =  
 i \bar \Psi(x) {\delta \Gamma \over \delta \bar \Psi(x)}
 + i  {\delta \Gamma \over \delta \Psi(x)} \Psi(x) .
 \label{WTL4}
\end{eqnarray}
The WT identity for the fermion-boson vertex function is
obtained by taking derivatives of both side of
Eq.~(\ref{WTL4}) in $\bar \Psi(z)$ and $\Psi(y)$ and
finally setting all the classical fields equal to zero, 
$J^\mu = \eta = \bar \eta = 0$:
\begin{eqnarray}
  {1 \over e} 
  \partial_\mu^x {\delta^3 \Gamma \over \delta A^\mu(x)
  \delta \Psi(y) \delta \bar \Psi(z)}  
 +  i{\delta^2 \Gamma \over 
 \delta \Psi(y) \delta \bar \Psi(z)}
 \delta(x-z)
 - i {\delta^2 \Gamma \over 
 \delta \Psi(y) \delta \bar\Psi(z)} 
 \delta(x-y)  =  0 .
 \label{WTL6}
\end{eqnarray}
 From the definitions:
\begin{eqnarray}
\Gamma_\mu(x,y,z)  
  &:=&  {\delta^3 \Gamma \over \delta A^\mu(x)
  \delta \Psi(y) \delta \bar \Psi(z)}  ,
\end{eqnarray}
and
\begin{eqnarray}
 {\delta^2 \Gamma \over \delta \Psi(y) \delta \bar \Psi(z)}
 = - \left[ {\delta^2 W \over \delta \eta(y) \delta
\bar \eta(z)} \right]^{-1}
 = \langle \Psi(y) \bar \Psi(x) \rangle^{-1} ,
\end{eqnarray}
we can see that (\ref{WTL6}) reproduces (\ref{WTL5}) in
momentum representation, using the Fourier transformation:
\begin{eqnarray}
   (2\pi)^D \delta^D(k-p+q) \Gamma_\mu(q,p) 
   &:=&  \int d^D x d^D y d^Dz 
   e^{i(q \cdot y - p \cdot z - k \cdot x)}
   \Gamma_\mu(x,y,z) .
\end{eqnarray}
The form (\ref{WTL6}) for the WT identity is known to be
very useful in the proof of renormalizability, which is
not the subject of this paper.

\section{Derivation of the transverse WT identity} 
\setcounter{equation}{0}

The usual WT identity for the vector vertex has been 
obtained by taking the divergence of the current
expectation value, Eq.~(\ref{WTL3}). 
Furthermore, we want to find the relation which specifies
the rotation of the current expectation value:
\begin{eqnarray}
 {\cal R}_{\mu\nu}(x)_J
 := \partial_\mu \langle {\cal J}_\nu(x) \rangle_J
  - \partial_\nu \langle {\cal J}_\mu(x) \rangle_J .
\end{eqnarray}
First of all, we find from Eq.~(\ref{WTA1}) and
Eq.~(\ref{WTA2})
\begin{eqnarray}
  {\cal R}_{\mu\nu}(x)_J
&=&  - {1 \over e} \left[
 \partial_\nu \Delta_{\mu\rho}(\partial) 
- \partial_\mu \Delta_{\nu\rho}(\partial)
 \right] A_\rho(x)
 - {1 \over e} [\partial_\nu J_\mu - \partial_\mu J_\nu ]
 \nonumber\\
&=&   - {1 \over e} \partial^2 \left[ 
 \partial_\mu {\delta W \over \delta J^\nu}
 - \partial_\nu {\delta W \over \delta J^\mu} \right]
 - {1 \over e} [\partial_\nu J_\mu - \partial_\mu J_\nu ]
\nonumber\\
&=& - {1 \over e}  \partial^2 
\left[\partial_\mu A_\nu(x)_J - \partial_\nu A_\mu(x)_J 
\right]  
 - {1 \over e} \left[
   \partial_\nu {\delta \Gamma \over \delta A^\mu(x)} 
 - \partial_\mu {\delta \Gamma \over \delta A^\nu(x)} 
 \right],
 \label{WTA3}
\end{eqnarray}
where we have used
\footnote{
In contrast to the longitudinal part, the transverse part
does not in general have the same form as the free case.
}
\begin{eqnarray}
 \partial_\nu \Delta_{\mu\rho}(\partial) 
- \partial_\mu \Delta_{\nu\rho}(\partial)
= \partial^2 
(\partial_\mu g_{\nu\rho} - \partial_\nu g_{\mu\rho}) .
\end{eqnarray}
\par
Next, we need to know another expression for the rotation
${\cal R}_{\mu\nu}^J$.
In order to derive such relations, it turns out that we
have only to pay attention to the the fermionic part 
\begin{eqnarray}
  {\cal L}_{F}  =   
   \bar \Psi i \gamma^\mu (\partial_\mu - ie A_\mu) \Psi
  - \bar \Psi M \Psi + \bar \eta \Psi + \bar \Psi \eta .
\end{eqnarray}
Therefore the following relations hold also for the
non-Abelian case irrespective of the gauge part ${\cal
L}_g$, if we identify $A_\mu$ in ${\cal L}_{F}$ as
$A_\mu(x)=A_\mu^a(x)T^a$ ($a=1, ..., dim G$) with the
generator $T^a$ of the gauge group $G$. 
\par
The identity
\begin{eqnarray}
  \int {\cal D}\bar \Psi {\cal D}\Psi {\cal D}A_\mu
  {\delta \over \delta \bar \Psi(x)}
  \exp \left[i \int d^D x({\cal L}+{\cal L}_{J}) 
  \right]
 \equiv 0,
 \label{ID2}
\end{eqnarray}
leads to
\begin{eqnarray}
   [[ i \gamma^\mu [
   \partial_\mu  - ie A_\mu(x) ]\Psi(x)
   - M \Psi(x) + \eta(x)    ]]_J = 0,
   \label{F0}
\end{eqnarray}
where we have omitted to write the indices explicitly.
We multiply Eq.(\ref{F0}) by the matrix $\Gamma$ from the
left where
$\Gamma$ may have spinor, flavor and color indices. And
then, operating the differential operator
${\delta  \over \delta \eta(y)}$ to the resulting equation,
we obtain
\begin{eqnarray}
  && \langle - \bar \Psi(y) i \Gamma \gamma^\mu [
   \partial_\mu - ie A_\mu(x)] \Psi(x)
   + \bar \Psi(y) \Gamma M \Psi(x)
   - \bar \Psi(y) \Gamma \eta(x) 
   \rangle_J 
   \nonumber\\
   &=& - {\rm tr}(\Gamma)\delta(x-y),
   \label{F1}
\end{eqnarray}
where we should be careful with the anticommuting nature of
the Grassmann variable and the ordering of spinor, flavor
and color indices, and the trace is taken over all the
indices.
\par
On the other hand, the identity
\begin{eqnarray}
  \int {\cal D}\bar \Psi {\cal D}\Psi {\cal D}A_\mu
  {\delta \over \delta \Psi(x)}
  \exp \left[i \int d^D x({\cal L}+{\cal L}_{J}) 
  \right]
 \equiv 0,
 \label{ID3}
\end{eqnarray}
leads to
\begin{eqnarray}
 [[ \bar \Psi(x) i \gamma^\mu [
 \overleftarrow{\partial_\mu} + ie A_\mu(x)]
   + m \bar \Psi(x) - \bar \eta(x)    ]]_J = 0 .
   \label{G0}
\end{eqnarray}
Similarly, multiplying Eq.~(\ref{G0}) by the same matrix
$\Gamma$ as above from the right and subsequently
operating 
${\delta \over \delta \bar \eta(y)}$, we obtain
\begin{eqnarray}
 && \langle - \bar \Psi(x) i [
 \overleftarrow{\partial_\mu} + ie A_\mu(x)] 
 \gamma^\mu  \Gamma \Psi(y)
   - \bar \Psi(x) M \Gamma \Psi(y) 
   + \bar \eta(x) \Gamma \Psi(y)
   \rangle_J 
   \nonumber\\
   &=& {\rm tr}(\Gamma) \delta^D(x-y) .
   \label{G1}
\end{eqnarray}
\par
By adding Eq.~(\ref{G1}) to Eq.~(\ref{F1}) or
subtracting  Eq.~(\ref{G1}) from Eq.~(\ref{F1}), and
subsequently setting $x=y$, i.e.  
$
 \left[ {\delta \over \delta \eta(y)} 
  [ \Gamma \times (\ref{F0})]  
 \pm {\delta \over \delta \bar \eta(y)} 
  [(\ref{G0}) \times \Gamma ] \right] \Big|_{x=y},
$
we get two sets of WT identities:
\begin{eqnarray}
 \partial_\rho \langle
 \bar \Psi {i \over 2} \{ \Gamma, \gamma^\rho \} \Psi 
 \rangle_J 
 &=& 
   \langle \bar \Psi [ \Gamma, M ] \Psi \rangle_J 
 - \langle \bar \Psi \Gamma \eta \rangle_J 
 + \langle \bar \eta \Gamma \Psi \rangle_J 
 \nonumber\\&&
 - \langle \bar \Psi {i \over 2}[\Gamma, \gamma^\rho] 
 (\overrightarrow{\partial_\rho} -
\overleftarrow{\partial_\rho}) \Psi \rangle_J 
+ ie \langle \bar \Psi i [\Gamma, \gamma^\rho A_\rho] \Psi 
 \rangle_J  
 \nonumber\\&&
 + {\cal A}_\Gamma[A_\mu],
\label{WTa}
\end{eqnarray}
\begin{eqnarray}
 \partial_\rho \langle
 \bar \Psi {i \over 2} [ \Gamma, \gamma^\rho ] \Psi 
 \rangle_J 
 &=& 
   \langle \bar \Psi \{ \Gamma, M \} \Psi \rangle_J 
 - \langle \bar \Psi \Gamma \eta \rangle_J 
 - \langle \bar \eta \Gamma \Psi \rangle_J 
 \nonumber\\&&
 - \langle \bar \Psi {i \over 2} \{ \Gamma, \gamma^\rho \} 
 (\overrightarrow{\partial_\rho} -
\overleftarrow{\partial_\rho}) \Psi \rangle_J  
+ ie \langle \bar \Psi i \{\Gamma, \gamma^\rho A_\rho\}
\Psi   \rangle_J 
 \nonumber\\&&
 + 2 {\rm tr}(\Gamma) \delta^D(0) 
 + {\cal A}_\Gamma[A_\mu],
\label{WTb}
\end{eqnarray}
where we have introduced the commutation relation 
$[A,B]:=AB-BA$ and
the anticommutation relation 
$\{A,B\}:=AB+BA$. 
Here ${\cal A}_\Gamma[A_\mu]$ denotes the possible anomaly
which will be discussed in section 5.

\par
In what follows, we choose $\Gamma$ to be a direct product
of the each factor for spinor, flavor and color indices:
$\Gamma = \Gamma_S \otimes \Gamma_F \otimes \Gamma_C$.
It turns out that the usual WT identity Eq.~(\ref{WTL0})
for the Abelian vector current is obtained from
Eq.~(\ref{WTa}) by choosing
\begin{eqnarray}
\Gamma = 1_S \otimes 1_F \otimes 1_C .
\end{eqnarray}

\par
A new type of WT identities, so-called the transverse WT
identity \cite{Takahashi86}, is obtained from
Eq.~(\ref{WTb}) by choosing 
\begin{eqnarray}
 \Gamma = \sigma_{\mu\nu} \otimes 1_F \otimes 1_C ,
\quad
 \sigma_{\mu\nu} := {i \over 2}[ \gamma_\mu, \gamma_\nu ] .
\end{eqnarray}
Indeed, this choice of $\Gamma$ yields
\begin{eqnarray}
 \partial_\rho \langle
 \bar \Psi(x) {i \over 2} [ \Gamma, \gamma^\rho ] \Psi(x)
 \rangle_J 
 &=& \partial_\mu 
 \langle \bar \Psi(x) \gamma_\nu \Psi(x) \rangle_J 
- \partial_\nu 
 \langle \bar \Psi(x) \gamma_\mu \Psi(x)  \rangle_J .
 \label{rotation}
\end{eqnarray}
Here it should be remarked that
\begin{eqnarray}
  \sigma_{\mu\nu} \gamma_\rho
  = {1 \over 2}[ \sigma_{\mu\nu}, \gamma_\rho ]
  + {1 \over 2}\{ \sigma_{\mu\nu}, \gamma_\rho \}
  = i(\gamma_\mu g_{\nu\rho} - \gamma_\nu g_{\mu\rho})
  + {1 \over 2}\{ \sigma_{\mu\nu}, \gamma_\rho \},
\end{eqnarray}
which follows from 
$
 \{ \gamma_\mu, \gamma_\nu \} = 2 g_{\mu\nu} .
$
(As we will see shortly, the symmetrized part 
$\{ \sigma_{\mu\nu}, \gamma_\rho \}$ changes depending on
the space-time dimension.)
Thus we arrive at the desired expression (up to anomaly):
\begin{eqnarray}
{\cal R}_{\mu\nu}(x)_J
&=&   
\langle \bar \Psi(x) \{ \sigma_{\mu\nu}, M \} \Psi(x)
\rangle_J 
 - \langle \bar \Psi(x) \sigma_{\mu\nu} \eta(x) \rangle_J 
 - \langle \bar \eta(x) \sigma_{\mu\nu} \Psi(x) \rangle_J 
 \nonumber\\&&
 - \langle \bar \Psi(x) 
 {i \over 2}\{ \sigma_{\mu\nu}, \gamma^\rho \}
 (\overrightarrow{\partial}_\rho  -
 \overleftarrow{\partial_\rho}) \Psi(x) \rangle_J  
 \nonumber\\&&
 -  e \langle \bar \Psi(x) 
 \{ \sigma_{\mu\nu}, \gamma_\rho \}
 \Psi(x) A^\rho(x) \rangle_J .
 \label{WTT1}
\end{eqnarray}
Taking derivatives of both side of Eq.~(\ref{WTT1}) with
respect to
$\bar \Psi(z)$ and $\Psi(y)$ and setting all the sources
equal to zero, 
$J^\mu = \eta = \bar \eta = 0$,
we get the transverse WT identity \cite{Takahashi86}:
\begin{eqnarray}
{\cal R}_{\mu\nu}(x,y,z) &:=&
\partial_\mu \langle \bar \Psi(x) \gamma_\nu \Psi(x) ;
\Psi(y) \bar \Psi(z) \rangle_c 
- \partial_\nu 
 \langle \bar \Psi(x) \gamma_\mu \Psi(x)  ;
\Psi(y) \bar \Psi(z) \rangle_c
 \nonumber\\ 
&=&   
\langle \bar \Psi(x) \{ \sigma_{\mu\nu}, M \} \Psi(x);
\Psi(y) \bar \Psi(z) \rangle_c
 \nonumber\\&&
 - \langle \Psi(y) \bar \Psi(x) 
\rangle_c \sigma_{\mu\nu} \delta^D(x-z)
 - \sigma_{\mu\nu} \langle \Psi(x) \bar
\Psi(z)  \rangle_c  \delta^D(x-y) 
 \nonumber\\&&
 - \langle \bar \Psi(x) 
 {i \over 2}\{ \sigma_{\mu\nu}, \gamma^\rho \}
 (\overrightarrow{\partial}_\rho  -
 \overleftarrow{\partial_\rho}) \Psi(x);
\Psi(y) \bar \Psi(z)  \rangle_c  
 \nonumber\\&&
 -  e \langle \bar \Psi(x) 
 \{ \sigma_{\mu\nu}, \gamma_\rho \}
 \Psi(x) A^\rho(x);
\Psi(y) \bar \Psi(z)  \rangle_c  .
 \label{WTT2}
\end{eqnarray}
The transverse WT identities were first found
by Takahashi based on the canonical formalism in the 3+1
dimension where
\begin{eqnarray}
 {1 \over 2}\{ \sigma_{\mu\nu}, \gamma_\rho \}
 = \epsilon_{\mu\nu\rho\sigma} \gamma_5 \gamma^\sigma .
\end{eqnarray}
In this paper we rederived them based on the path
integral formalism.  We find that the transverse WT
identity exhibits different  appearance depending on the
dimensionality of space-time. 
\par
In 2+1 dimensions, we choose the gamma
matrices as
\begin{eqnarray}
\gamma^0 = \sigma_3, \quad
\gamma^1 = i \sigma_1, \quad
\gamma^2 = i \sigma_2 .
\end{eqnarray}
Then 
$
\sigma_{\mu\nu}=- \epsilon_{\mu\nu\sigma}
\gamma^\sigma (\epsilon_{012}=1)
$ 
and hence
\begin{eqnarray}
 {1 \over 2}\{ \sigma_{\mu\nu}, \gamma_\rho \}
 = - \epsilon_{\mu\nu\rho}  .
\end{eqnarray}
\par
And in 1+1 dimensional space-time, we choose
\begin{eqnarray}
\gamma^0 = \sigma_2, \quad
\gamma^1 =  i \sigma_1, \quad
\gamma^5 := \gamma^0 \gamma^1 =  \sigma_3,\quad
(\epsilon_{01} = 1),
\end{eqnarray}
which implies 
\begin{eqnarray}
 \sigma_{\mu\nu} = i \epsilon_{\mu\nu} \gamma_5 ,
\quad
\gamma_\mu \gamma_5 = \epsilon_{\mu\nu}\gamma^\nu .
\end{eqnarray}
Therefore, we obtain
\begin{eqnarray}
  \{ \sigma_{\mu\nu}, \gamma_\rho \}
 = i\epsilon_{\mu\nu} \{ \gamma_5, \gamma_\rho \}
 \equiv 0 .
\end{eqnarray}
In 1+1 dimensions, therefore, Eq.~(\ref{WTT1})  and
Eq.~(\ref{WTT2}) have remarkably simple forms:
\begin{eqnarray}
{\cal R}_{\mu\nu}(x)_J
=   
\langle \bar \Psi(x) \{ \sigma_{\mu\nu}, M \} \Psi(x)
\rangle_J 
 - \langle \bar \Psi(x) \sigma_{\mu\nu} \eta(x) \rangle_J 
 - \langle \bar \eta(x) \sigma_{\mu\nu} \Psi(x) \rangle_J ,
\end{eqnarray}
\begin{eqnarray}
{\cal R}_{\mu\nu}(x,y,z) 
&=&  
\langle \bar \Psi(x) \{ \sigma_{\mu\nu}, M \} \Psi(x);
\Psi(y) \bar \Psi(z) \rangle_c
 \nonumber\\&&
 - \langle \Psi(y) \bar \Psi(x) 
\rangle_c \sigma_{\mu\nu} \delta^D(x-z)
 - \sigma_{\mu\nu} \langle \Psi(x) \bar
\Psi(z)  \rangle_c  \delta^D(x-y) 
 \label{WTT3}
\end{eqnarray}
In the chiral limit $M=0$, especially, the transverse
WT identity leads to the surprisingly simple identity
for the rotation of the vector vertex.
For example, in QED$_2$, we obtain by making use of
Eq.~(\ref{FT1}):
\begin{eqnarray}
  k_\mu \Gamma_\nu(q,p) - k_\nu \Gamma_\mu(q,p)
  = S^{-1}(q)\sigma_{\mu\nu} + \sigma_{\mu\nu}S^{-1}(p) .
 \label{WTT4}
\end{eqnarray}

\par
The non-Abelian versions of the transverse WT identities as
well as the usual longitudinal WT identities  for the
current 
\begin{eqnarray}
  {\cal J}_\mu^a(x) := \bar \Psi(x) \gamma_\mu T^a \Psi(x),
\end{eqnarray}
are obtained
from (\ref{WTa}) and (\ref{WTb}), if we set 
$\Gamma = \sigma_{\mu\nu} \otimes 1_F \otimes T^a$ and
$\Gamma = 1 \otimes 1_F \otimes T^a$ respectively:
the longitudinal WT identity reads
\begin{eqnarray}
  \langle D_\mu[A]^{ab} {\cal J}^\mu{}^b(x) \rangle_J
 = i \langle \bar \Psi(x) T^a \eta(x) \rangle_J
 - i \langle \bar \eta(x) T^a \Psi(x) \rangle_J,
 \label{WTLna}
\end{eqnarray}
while the transverse WT identity is given by
\begin{eqnarray}
 && \langle D_\mu[A] {\cal J}_\nu(x) \rangle_J
- \langle D_\nu[A] {\cal J}_\mu(x) \rangle_J
\nonumber\\
&=&   
2M \langle \bar \Psi(x) \sigma_{\mu\nu} T^a \Psi(x)
\rangle_J 
 - \langle \bar \Psi(x) \sigma_{\mu\nu} T^a \eta(x)
\rangle_J 
 - \langle \bar \eta(x) \sigma_{\mu\nu} T^a \Psi(x)
\rangle_J 
 \nonumber\\&&
 - \langle \bar \Psi(x) 
 {i \over 2}\{ \sigma_{\mu\nu}, \gamma^\rho \} T^a
 (\overrightarrow{\partial_\rho}  -
 \overleftarrow{\partial_\rho}) \Psi(x) \rangle_J  
 \nonumber\\&&
 -  e \langle \bar \Psi(x) {1 \over 2}
 \{ \sigma_{\mu\nu}, \gamma_\rho \} \{ T^a, T^b \}
 \Psi(x) A^\rho{}^b(x) \rangle_J  ,
 \label{WTTna}
\end{eqnarray}
where the covariant derivative $D_\mu[A]$ is defined by
\begin{eqnarray}
   D_\mu[A]^{ab}
   : = \delta^{ab} \partial_\mu  + e f^{abc} A_\mu^b ,
 \label{cd}
\end{eqnarray}
for our convention $[T^a, T^b] = if^{abc}T^c$.

\section{Chiral WT identity} 
\setcounter{equation}{0}
As a special case of the identities, (\ref{WTa}) and
(\ref{WTb}), we can derive the chiral WT identities, i.e.
WT identity for the axial vector current
\begin{eqnarray}
{\cal J}_\mu^5(x) 
:= \bar \Psi(x) \gamma_5 \gamma_\mu \Psi(x) .
\end{eqnarray}
The choice 
\begin{eqnarray}
\Gamma = \gamma_5 \otimes 1_F \otimes 1_C
\end{eqnarray}
in Eq.~(\ref{WTb}) yields the usual chiral WT identity:
\begin{eqnarray}
 \partial_\mu \langle {\cal J}^\mu_5(x) \rangle_J
 =  2M \langle \bar \Psi(x) \gamma_5 \Psi(x) \rangle_J
 - \langle \bar \Psi(x) \gamma_5 \eta(x) \rangle_J
 - \langle \bar \eta(x) \gamma_5 \Psi(x) \rangle_J,
 \label{cWTL0}
\end{eqnarray}
up to the quantum anomaly ${\cal A}[A_\mu]$ which will be
discussed in the next section.
Up to the anomalous term, this leads to
\begin{eqnarray}
{\cal D}_5(x,y,z) 
&:=& \partial_\mu \langle {\cal J}^\mu_5(x); 
 \Psi(y); \bar \Psi(z) \rangle_c
 \nonumber\\
 &=&  \langle \bar \Psi(x) \{ \gamma_5, M \} \Psi(x) 
 \Psi(y) \bar \Psi(z) \rangle_c
 \nonumber\\&&
-  \langle \Psi(y) \bar \Psi(z) \rangle_c 
    \gamma_5 \delta(x-z)
 -  \gamma_5 
 \langle \Psi(y) \bar \Psi(z) \rangle_c \delta(x-y) 
 \nonumber\\&&
 + \langle {\cal A}[A_\mu]; 
 \Psi(y); \bar \Psi(z) \rangle_c .
\end{eqnarray}
Moreover, choosing
\begin{eqnarray}
 \Gamma = \gamma_5 \sigma_{\mu\nu}
 \otimes 1_F \otimes 1_C
\end{eqnarray}
in Eq.~(\ref{WTa}), we obtain the transverse WT identity for
the axial vector vertex:
\begin{eqnarray}
{\cal R}_{\mu\nu}^5(x)_J
&:=&   
 - \langle \bar \Psi(x) \gamma_5\sigma_{\mu\nu} \eta(x) 
 \rangle_J
 - \langle \bar \eta(x) \sigma_{\mu\nu}\gamma_5 \Psi(x) 
 \rangle_J
 \nonumber\\&&
 - \langle \bar \Psi(x) {1 \over 2}
 \{ \sigma_{\mu\nu}, \gamma^\rho \} \gamma_5
 (\overrightarrow{\partial_\rho}  -
 \overleftarrow{\partial_\rho}) \Psi(x) \rangle_J  
 \nonumber\\&&
 -  e \langle \bar \Psi(x) 
 \{ \sigma_{\mu\nu}, \gamma_\rho \} \gamma_5
 \Psi(x) A^\rho(x) \rangle_J  .
 \label{cWTT1}
\end{eqnarray}
Then we obtain
\begin{eqnarray}
{\cal R}_{\mu\nu}^5(x,y,z)
&:=& \partial_\mu \langle {\cal J}_\nu^5(x) ;
 \Psi(y); \bar \Psi(z) \rangle_c
- \partial_\nu \langle {\cal J}_\mu^5(x) ;
 \Psi(y); \bar \Psi(z) \rangle_c
 \nonumber\\ 
 &=& 
 - \langle \Psi(y) \bar \Psi(x) 
\rangle_c \gamma_5 \sigma_{\mu\nu} \delta^D(x-z)
 -  \sigma_{\mu\nu} \gamma_5 \langle \Psi(x) \bar
\Psi(z)  \rangle_c  \delta^D(x-y) 
 \nonumber\\&&
 - \langle \bar \Psi(x) {1 \over 2}
 \{ \sigma_{\mu\nu}, \gamma^\rho \} \gamma_5
 (\overrightarrow{\partial_\rho}  -
 \overleftarrow{\partial_\rho}) 
 \Psi(x); \Psi(y); \bar \Psi(z) \rangle_c  
 \nonumber\\&&
 - e \langle \bar \Psi(x) 
 \{ \sigma_{\mu\nu}, \gamma_\rho \} \gamma_5
 \Psi(x) A^\rho(x); \Psi(y); \bar \Psi(z) \rangle_c .
\end{eqnarray}
In 1 + 1 dimensional case, the transverse chiral WT
identity reduces to the longitudinal (vector) WT identity:
\begin{eqnarray}
{\cal R}_{\mu\nu}^5(x)_J
= \epsilon_{\mu\nu}{\cal D}(x)_J,
\quad
{\cal R}_{\mu\nu}^5(x,y,z)
= \epsilon_{\mu\nu}{\cal D}(x,y,z) ,  
\end{eqnarray}
since
$
 {\cal J}_\mu^5 = \epsilon_{\mu\nu}{\cal J}^\nu ,
$
and hence
$
 \partial_\mu {\cal J}_\nu^5 
 - \partial_\nu {\cal J}_\mu^5 
 = \epsilon_{\mu\nu} \partial^\rho {\cal J}_\rho .
$
On the other hand, the longitudinal chiral WT identity
reduces to the transverse (vector) WT identity:
\begin{eqnarray}
{\cal D}_5(x)_J 
= \epsilon^{\mu\nu} {\cal R}_{\mu\nu}^5(x)_J ,
\quad
{\cal D}_5(x,y,z)
= \epsilon^{\mu\nu} {\cal R}_{\mu\nu}^5(x,y,z) ,
\end{eqnarray}
since
$
 \partial^\mu {\cal J}_\mu^5 
 = \epsilon_{\mu\nu} \partial^\mu {\cal J}^\nu
 = {1 \over 2} \epsilon_{\mu\nu} 
 (\partial^\mu {\cal J}^\nu - \partial^\nu {\cal J}^\mu) .
$
The extension to the non-Abelian case is straightforward
for the axial vector current
\begin{eqnarray}
  {\cal J}_\mu^5{}^a(x) 
  := \bar \Psi (x)  \gamma_5 \gamma_\mu T^a \Psi(x),
\end{eqnarray}
as in the previous section.

\section{Anomaly} 
\setcounter{equation}{0}

\subsection{general case}

In the derivation of the WT identity so far, we have not
taken into account the anomaly.  
Now we examine a possibility of the existence of the
anomaly based on the path integral formalism.  
The method of deriving the anomaly in the path integral
formalism is well known as Fujikawa's method
\cite{Fujikawa79}.
The anomaly ${\cal A}[A](x)$ comes from the Jacobian factor
$J[A]$ which is accompanied by the change of variable
in the path integral measure:
\footnote{
The Fujikawa method is applied only in the Euclidean
space.  Hence the anomaly obtained in this method is the
Euclidean version of the corresponding anomaly in
Minkowski space-time, although we write the anomaly as if
we obtained it in Minkowski space-time in what
follows. }
\begin{eqnarray}
 {\cal D}\bar \Psi' {\cal D} \Psi' 
 = {\cal D}\bar \Psi  {\cal D} \Psi  J[A],
 \quad
 J[A] = \exp \left[ - \int d^Dx \alpha(x) {\cal A}[A](x)
\right].
\end{eqnarray}
Actually, all the WT identities derived so far in this paper
can also be derived by appropriate change of variable
together with the possible anomaly, as shown in the
following. In the calculation of the anomaly, the
gauge-boson field is identified as the external field. 
Therefore, it is enough to pay attention only to the
fermionic part ${\cal L}_{F}$. Perform the following change
of variable:
\footnote{
In the Euclidean space, $\Psi$ and $\bar \Psi$ are
independent Grassmann variables, in sharp contrast with the
Minkowski space-time where 
$\bar \Psi = \Psi^\dagger \gamma^0$.
Therefore, the independent change of variable is allowed
in Euclidean space.  
}
\begin{eqnarray}
  \Psi(x) &\rightarrow& 
  \Psi'(x) := e^{i \Omega \alpha(x)} \Psi(x),
\nonumber\\
\bar \Psi (x) &\rightarrow& \bar \Psi '(x) 
 := \bar \Psi(x) e^{i \tilde \Omega\alpha(x)},
 \label{cov}
\end{eqnarray}
where $\Omega$ and $\tilde \Omega$ are global quantities
which are similar to $\Gamma$ used in the  previous
derivation of the WT identity.
If we put $\tilde \Omega = \pm \Omega$, the change of the
fermionic action
$S_F=\int d^D x {\cal L}_{F}$ under the transformation
(\ref{cov}) is written (for small parameter $\alpha$) as
\begin{eqnarray}
  \delta_\Omega S_{F}  &=&   \int d^D x \Big\{
 \bar \Psi {1 \over 2}[ i \gamma^\mu, 
 i \Omega \partial_\mu \alpha ]_{\mp} \Psi
 \nonumber\\&&
+ \bar \Psi {1 \over 2}[ \gamma^\mu, 
 i \Omega \alpha ]_{\pm} \partial_\mu \Psi
 - \partial_\mu \bar \Psi {1 \over 2}[ \gamma^\mu, 
 i \Omega \alpha ]_{\pm} \Psi
 \nonumber\\&&
+ e \bar \Psi [ \gamma^\mu, 
 i \Omega \alpha ]_{\pm} \Psi
-  \bar \Psi [ M ,  i \Omega \alpha ]_{\pm} \Psi 
 \nonumber\\&&
+ \bar \eta i \Omega \alpha \Psi
+ \bar \Psi i \tilde \Omega \alpha \eta \Big\}
+ O(\alpha^2),
\end{eqnarray}
where we have employed the notation for the anticommutation
and the commutation relations:
$[A,B]_{+}:=\{A,B\}$ and $[A,B]_{-}:=[A,B]$.
The theory is invariant under the change of variable:
\begin{eqnarray}
 0 =   \langle {\delta_\Omega S \over \delta
\alpha^a(x)}\Biggr|_{\alpha=0} \rangle_J .
\end{eqnarray}
Then, after performing the integration by parts, we obtain
\begin{eqnarray}
&& \partial_\rho \langle \bar \Psi(x) {1 \over 2}[ i
\gamma^\rho, 
 i \Omega^a  ]_{\mp} \Psi(x) \rangle_J
 \nonumber\\ 
&=& 
-  \langle  \bar \Psi(x) [ M ,  i \Omega^a ]_{\pm}
\Psi(x) \rangle_J 
+ \langle \bar \eta(x) i \Omega^a \Psi(x) \rangle_J
\pm \langle \bar \Psi(x) i \Omega^a \eta(x) \rangle_J
 \nonumber\\&&
+ \langle \bar \Psi(x) {1 \over 2}[ \gamma^\rho, 
 i \Omega^a ]_{\pm} (\overrightarrow{\partial_\rho}  -
 \overleftarrow{\partial_\rho}) \Psi(x)  \rangle_J
 \nonumber\\&&
+ e \langle \bar \Psi(x) [ \gamma^\rho A_\rho(x), 
 i \Omega^a ]_{\pm} \Psi(x)\rangle_J
+ \langle {\cal A}_\Omega[A](x) \rangle_J ,
\label{WTc}
\end{eqnarray}
where we have set $\alpha(x)=\alpha^a(x)T^a$ $(a= 1, ...,
dim G)$ and defined $\Omega^a:=\Omega T^a$.  
In the Abelian case, we can omit the index and 
put $T^a \equiv 1$. 
Here ${\cal A}_\Omega[A](x)$ denotes symbolically the
anomaly accompanied by the above transformation
(\ref{cov}). Up to the anomaly, (\ref{WTc}) is
equivalent to a pair of Eq.~(\ref{WTa}) and
(\ref{WTb}). 

\par
First, we consider the case $\tilde \Omega = + \Omega$ which
corresponds to the upper signature in Eq.~(\ref{WTc}). 
In this case,
\begin{eqnarray}
\Omega= \gamma_5 \otimes 1_F \otimes 1_C
\end{eqnarray}
reproduces the chiral WT identity.
Actually, the change of variable (\ref{cov}) reproduces
correctly the chiral anomaly as shown by Fujikawa
\cite{Fujikawa79}.  Especially, for Abelian gauge
theory in $D=4$ dimensions \cite{ABJ69}
\begin{eqnarray}
 {\cal A}_\Omega[A] 
 = {-i \over 16\pi^2}  \epsilon^{\mu\nu\alpha\beta} 
 \ F_{\mu\nu}F_{\alpha\beta} .
\end{eqnarray}
The transverse WT identity is reproduced for
\begin{eqnarray}
\Omega= \epsilon^{\mu \nu} \sigma_{\mu\nu} \otimes
1_F \otimes 1_C .
\end{eqnarray}
In this case, the change of variable (\ref{cov}) agrees
with the local Lorentz transformation
\begin{eqnarray}
\Omega \alpha(x) 
= \epsilon^{\mu \nu} \sigma_{\mu\nu} \alpha(x).
\end{eqnarray}
Usually it is understood that there is no Lorentz
anomaly in the theory of the Dirac fermion in the flat
space-time, in contrast with the parity-breaking Weyl
fermion in the presence of non-trivial background
gravitational field in $ D \ge 4$ dimensions
\cite{ganomaly}. Therefore we conclude that there is no
anomaly for the transverse WT identity and hence there is no
correction to the expression in $D \ge 4$ dimensions for the
transverse WT identity (\ref{WTT1}) derived in the previous
method.
\par
On the other hand, the case of
$
\tilde \Omega = - \Omega
$
corresponds to the lower
signature in Eq.~(\ref{WTc}). 
The longitudinal (vector) WT identity is obtained from
\begin{eqnarray}
\Omega = 1_S \otimes 1_F \otimes 1_C .
\end{eqnarray} 
Finally, the transverse chiral WT identity corresponds to
\begin{eqnarray}
\Omega = \epsilon^{\mu \nu} \gamma_5 \sigma_{\mu\nu} \otimes
1_F \otimes 1_C .
\end{eqnarray} 
For this choice, as far as we know, there are no
references which have investigated the corresponding
anomaly.
As we do not need the detail of this case in the following
part of this article and its investigation is
off the main stream of this article, we do not pursue this
issue any longer. However, the 1+1 dimensional case is
simple and discussed below.  The higher dimensional case
deserves further studies.

\subsection{1+1 dimensions}

\par
It should be remarked that in 1+1 dimension the situation
is somewhat different from the higher dimensional case.
The transverse chiral WT identity is essentially the
same as the (longitudinal) WT identity, since
$
\Omega= \epsilon^{\mu \nu} \gamma_5 \sigma_{\mu\nu}
= \epsilon^{\mu \nu} \gamma_5 \epsilon_{\mu \nu} \gamma_5
=  2 1_S  .
$
They are both
anomaly free, if we impose the gauge invariance. 
Similarly, as
$
\Omega= \epsilon^{\mu \nu} \sigma_{\mu\nu} 
=  2 \gamma_5 ,
$
the transverse WT identity gives the equivalent
information to the chiral (longitudinal) WT identity in 1+1
dimensions. In this case, the information of the anomaly
can be also translated into the rotation of the
vector current, since the equation
\begin{eqnarray}
  \partial^\mu {\cal J}_\mu^5  
  =  {\cal A}_2,
  \quad
  {\cal A}_2 :=
  - {e \over 2\pi} \epsilon_{\mu\nu} F^{\mu\nu},
  \label{anomalyd=2}
\end{eqnarray}
is equivalent to 
\begin{eqnarray}
  \partial_\mu {\cal J}_\nu  - \partial_\nu {\cal J}_\mu 
  =  - {e \over \pi} F_{\mu\nu} .
  \label{anomalyd=2'}
\end{eqnarray}
Nevertheless we show that there is no modification in the
chiral WT identity (\ref{cWTT1}) due to the existence of
the chiral anomaly ${\cal A}_2$.
This is because the chiral anomaly is promoted to the
modification of the gauge-boson propagator as dynamical
mass generation (by radiative quantum correction)
$\mu: = e/\sqrt{\pi}$ for the gauge-boson $A_\mu$.  
In fact, for the vacuum polarization tensor in 1+1
dimensions of the form:
\begin{eqnarray}
 \Pi_{\mu\nu}(k) = {e^2 \over \pi} 
 (g_{\mu\nu} - {k_\mu k_\nu \over k^2} ) ,
 \label{vpd=2}
\end{eqnarray}
Eq.~(\ref{WTA3}) is modified as follows.
\begin{eqnarray}
&& -{1 \over e}[\partial_\nu D^{-1}_{\mu\rho}(\partial) 
- \partial_\mu D^{-1}_{\nu\rho}(\partial)] A^\rho
\nonumber\\ 
&=& -{1 \over e} (\partial^2 + {e^2 \over \pi})
(\partial_\mu g_{\nu\rho} - \partial_\nu g_{\mu\rho})
A^\rho
\nonumber\\ 
&=& -{1 \over e} [\partial_\nu
D^{(0)}{}^{-1}_{\mu\rho}(\partial)  - \partial_\mu
D^{(0)}{}^{-1}_{\nu\rho}(\partial)] A^\rho 
- {e \over \pi} F_{\mu\nu} .
\end{eqnarray}
Indeed, the additional term in this equation agrees with the
right-hand-side of Eq.~(\ref{anomalyd=2'}).

\section{Application to the SD equation} 
\setcounter{equation}{0}

In what follows we restrict our study to the Abelian gauge
theory.
\par
In momentum representation, the SD equation for the full
fermion propagator $S(p)$ is given by
\begin{eqnarray}
 S_0^{-1}(p)S(p) = 1 + ie^2  \int {d^Dk \over (2\pi)^D}
   \gamma^\mu D_{\mu\nu}(k) S(p-k) \Gamma_\nu(p-k,p) S(p),
   \label{SDf}
\end{eqnarray}
where $D_{\mu\nu}(k)$ is the full gauge-boson propagator,
$\Gamma_\nu(p-k,p)$ the full vertex function 
and $S_0$ the bare fermion propagator:
\begin{eqnarray}
 S_0(p) := {1 \over \hat{p}-m},
 \quad
 (\hat{p} := \gamma^\mu p_\mu) .
\end{eqnarray}
\par
Now we introduce more general gauge fixing than the
usual one, which is called the nonlocal gauge-fixing, see
e.g.
\cite{KM95}. In the configuration space, the gauge fixing
term in the nonlocal gauge is given by
\begin{equation}
  {\cal L}_{GF}
  = - {1 \over 2} F[A(x)] \int d^D y
  {1 \over \xi(x-y)}F[A(y)],
\label{nlgfterm}
\end{equation}
for the covariant gauge
$F[A]=\partial^\mu A_\mu$. 
In momentum representation, the gauge-fixing parameter
$\xi$ gets momentum-dependent, i.e.
$\xi$ becomes a function of the momentum:
$\xi=\xi(k^2)$.
Here it should be noted that  $\xi^{-1}(k^2)$ is the
Fourier transform of $\xi^{-1}(x)$, 
\begin{equation}
  \xi^{-1}(x) = \int {d^Dk \over (2\pi)^D}
e^{ikx} \xi^{-1}(k^2),
\quad
  \xi^{-1}(k^2) = \int  d^D x
e^{-ikx} \xi^{-1}(x),
\end{equation}
while $\xi(k^2)$ is not the Fourier
transform of $\xi(x)$, see ref.~\cite{KM95}.
If $\xi(k^2)$ does not have the
momentum-dependence, i.e.,
$\xi(k^2) \rightarrow \xi$, then 
$\xi^{-1}(x-y) \rightarrow \delta(x-y)\xi^{-1}$ and the
nonlocal gauge-fixing term reduces to the usual
gauge-fixing term: 
\begin{equation}
  {\cal L}_{GF}
  = - {1 \over 2\xi} (F[A(x)])^2 .
\label{gfterm}
\end{equation}
Hence, in the nonlocal gauge, the bare
gauge-boson propagator 
$D^{(0)}_{\mu\nu}(k) $ is given by
\begin{eqnarray}
 D^{(0)}_{\mu\nu}{}^{-1}(k) 
 = k^2 g_{\mu\nu} -  k_\mu k_\nu +  \xi(k)^{-1} k_\mu
k_\nu .
\end{eqnarray}
\par
On the other hand, the SD equation for the full gauge-boson
propagator is given by
\begin{eqnarray}
 D_{\mu\nu}^{-1}(k) 
 &=& D^{(0)}_{\mu\nu}{}^{-1}(k) - \Pi_{\mu\nu}(k),
 \nonumber\\
 \Pi_{\mu\nu}(k) &:=& e^2 \int {d^Dp \over (2\pi)^D}
 {\rm tr}[\gamma_\mu S(p) \Gamma_\nu(p,p-k) S(p-k)].
\end{eqnarray}
In the gauge theory, the vacuum polarization tensor
should have  the transverse form:
\begin{eqnarray}
 \Pi_{\mu\nu}(k)  
 =  \left( g_{\mu\nu} - {k_\mu k_\nu \over k^2} \right) 
 \Pi(k),
\end{eqnarray}
as far as the gauge invariance is preserved.
Hence the full gauge-boson propagator is of the form
\begin{eqnarray}
 D_{\mu\nu}(k) 
 &=& D_T(k) \left( g_{\mu\nu} - {k_\mu k_\nu \over k^2}
\right)
 + {\xi(k) \over k^2} {k_\mu k_\nu \over k^2}
 = D_{\mu\nu}^T(k) + D_{\mu\nu}^L(k) ,
 \nonumber\\
 D_T(k) &:=& {1 \over k^2 - \Pi(k)} .
\end{eqnarray}
\par
In the following we wish to given comments on how to choose
the ansatz for the vertex function $\Gamma_\mu$ in solving
the SD equation for the fermion propagator.

\subsection{decomposition of the vertex function}

To solve the SD equation for the fermion propagator, we
do not need to know the explicit form of the vertex function
$\Gamma_\mu$ itself, although much effort has been
devoted to find an ansatz for the full vertex function
\cite{Atkinson94,Pennington95,Roberts94,Pennington96}.   
It is enough to specify the divergence 
$\partial_\mu \Gamma^\mu$ and the rotation 
$\partial_\mu \Gamma_\nu - \partial_\nu \Gamma_\mu$
of the vertex, not the vertex itself, since the vertex
function appears only in this combination in the integrand
of the SD equation for the fermion propagator.  This can be
observed as follows.  In Eq.~(\ref{SDf}),
\begin{eqnarray}
&& D_{\mu\nu}(k) \tilde \Gamma^\nu(q,p)
 \nonumber\\
 &=& D_{\mu\nu}^L(k) \tilde \Gamma^\nu(q,p) 
 + D_{\mu\nu}^T(k) \tilde \Gamma^\nu(q,p)  
 \nonumber\\
 &=& {\xi(k^2) \over k^2}{k_\mu \over k^2}
 [k_\nu \tilde \Gamma^\nu(q,p)]
 +  D_T(k) {k^\nu \over k^2} 
[k_\nu \tilde \Gamma_\mu(q,p) 
- k_\mu \tilde \Gamma_\nu(q,p)],
\label{integrand}
\end{eqnarray}
where we have defined $q := p-k$ and
\begin{eqnarray}
 \tilde \Gamma_\nu(q,p) := S(q) \Gamma_\nu(q,p) S(p) .
\end{eqnarray}
The divergence 
$k_\nu \tilde \Gamma^\nu(q,p)$ is known exactly from the
usual WT identity (\ref{WTL5}).  For the rotation, 
$
k_\nu \tilde \Gamma_\mu(q,p) - k_\mu \tilde
\Gamma_\nu(q,p) , 
$
the transverse WT identity (\ref{WTT1}) gives a clue to find
the correct expression.  
Moreover, the combination
$\tilde \Gamma_\mu := S(p)\Gamma_\mu(p,q)S(q)$ is more
convenient rather than
$\Gamma_\mu(p,q)$ to specifying the vertex as shown in the
following subsections.
\par
The decomposition in (\ref{integrand}) can be written as 
\begin{eqnarray}
 D_{\mu\nu}(k) \tilde \Gamma^\nu(q,p)
= {\xi(k^2) \over k^2} 
 \tilde \Gamma_\mu{}^L(q,p)
 +  D_T(k) \tilde \Gamma_\mu{}^T(q,p) .
\end{eqnarray}
Therefore, the specification of the transverse vertex to the
form
\begin{eqnarray}
\Gamma_\mu{}^T(q,p) = 
{k^\nu \over k^2} 
[k_\nu \Gamma_\mu(q,p) 
- k_\mu \Gamma_\nu(q,p)]
= \left( g_{\mu\nu} - {k_\mu k_\nu \over k^2} \right)
\Gamma^\nu(q,p)
\end{eqnarray}
is {\it allowed only in the integrand of the SD equation
for the fermion propagator}, as well as 
\begin{eqnarray}
\Gamma_\mu{}^L(q,p) = 
{k^\mu \over k^2} 
[k_\nu \Gamma^\nu(q.p)]
= {k^\mu \over k^2} [S^{-1}(q) - S^{-1}(p)].
\end{eqnarray}
Note that the factor $1/k^2$ comes from the gauge-boson
propagator in the integrand, see \cite{Kondo92}

\subsection{gauge choice and LK transformation}

If we know a set of solutions (for the full gauge boson
propagator, the full fermion propagator and the full vertex
function) in a {\it single} gauge, the solutions in other
gauges are obtained through the Landau-Khalatnikov (LK)
transformation \cite{LK56}:
\begin{eqnarray}
  D_{\mu\nu}'(x) &=& D_{\mu\nu}(x) 
  + \partial_\mu  \partial_\nu 
  f(x),
  \nonumber\\
  S'(x,y) &=& e^{ e^2[f(o)-f(x-y)] } S(x,y),
  \nonumber\\
  {\cal V}_\nu'(x,y,z) 
  &=& e^{ e^2[f(o)-f(x-y)] }{\cal V}_\nu(x,y,z)
  \nonumber\\&&
  + S(x,y) e^{ e^2[f(o)-f(x-y)] } 
  \partial_\nu^z [f(x-z) - f(z-y)],
\label{LKtransf}
\end{eqnarray}
where
\begin{eqnarray}
D_{\mu\nu}(x,y) &=& \langle 0|T[A_\mu(x) A_\nu(y)]|0
\rangle,
  \nonumber\\
  S(x,y) &=& \langle 0|T[\Psi(x) \bar \Psi(y)]|0 \rangle ,
  \nonumber\\
  {\cal V}_\nu(x,y,z) 
 &=& \langle 0|T[\Psi(x) \bar \Psi(y) A_\nu(z)]|0 \rangle .
\end{eqnarray}
Here the function $f(x)$ allows a nonlocal gauge fixing.
Actually the SD equation is form-invariant under the LK
transformation. 
This can be easily shown in the coordinate space where the
SD equation has the following form:
\begin{eqnarray}
 (i \hat{\partial} - m) S(x,y) 
 = \delta^D(x-y) + i e^2 \gamma^\mu 
 \langle \Psi(x) \bar \Psi(y) A_\mu(x) \rangle,
\label{cSDf}
\end{eqnarray}
and
\begin{eqnarray}
D_{\mu\nu}^{-1}(x,z)  
&=&  D_{\mu\nu}^{(0)}{}^{-1}(x,z) - \Pi_{\mu\nu}(x,z) ,
  \nonumber\\
  \Pi_{\mu\nu}(x,z) 
  &=& (g_{\mu\nu}\partial^2 - \partial_\mu \partial_\nu) 
  \Pi(x-z)
  \nonumber\\
  &=& ie^2 \int d^Dz_1 d^Dz_2 {\rm tr}[\gamma_\mu
  S(x,z_1) \Gamma_\nu(z_1,z_2;z) S(z_2,x)],
\label{cSDg}
\end{eqnarray}
where 
\begin{eqnarray}
 \langle \Psi(x) \bar \Psi(y) A_\mu(z) \rangle
 = \int d^Dx' d^Dy' d^Dz' S(x,x') \Gamma_\nu(x',y';z')
S(y',y) D_{\mu\nu}(z',z) .
\label{vertex}
\end{eqnarray}

\par
Under the LK transformation, the divergence 
$\partial_\mu \tilde \Gamma^\mu$ and the rotation 
$\partial_\mu \tilde \Gamma_\nu - \partial_\nu \tilde
\Gamma_\mu$ 
formed from $\tilde \Gamma^\mu$, not from $\Gamma^\mu$, obey the following simple transformation law
which is the same as that of the full fermion propagator
(\ref{LKtransf}):
\begin{eqnarray}
  {\cal D}'(x,y,z) 
  &=& e^{ e^2[f(o)-f(x-y)] } {\cal D}(x,y,z),
  \nonumber\\
  {\cal R}_{\mu\nu}'(x,y,z) 
  &=& e^{ e^2[f(o)-f(x-y)] } {\cal R}_{\mu\nu}(x,y,z) .
\label{LKtransf2}
\end{eqnarray}
So the problem reduces to find a set of solutions in a
single gauge.

\par
\subsection{non-linearity}

As is usually done, if one starts from the SD equation for
the inverse fermion propagator
$S^{-1}(p)$:
\begin{eqnarray}
 S^{-1}(p) = S_0^{-1}(p) + ie^2  \int {d^Dk \over (2\pi)^D}
   \gamma^\mu D_{\mu\nu}(k) S(p-k) \Gamma_\nu(p-k,p) ,
   \label{SDf2}
\end{eqnarray}
and decomposes it into a pair of integral equations for
$A$ and  $B$ using
\begin{eqnarray}
 S^{-1}(p) = A(p)\hat{p}-B(p) ,
\end{eqnarray}
the SD equations become a pair of non-linear
integral equations for $A$ and $B$. Therefore, when one
wishes to solve the equation analytically, some type of
linearization is indispensable in most cases.  The
linearization approximation is sufficient to study some
restricted kinds of problems, e.g. the critical behavior in
the neighborhood of the critical point at which $B(p) \equiv
0$. However, the solution far from the critical point 
or the solution in the time-like region 
\cite{FK76,AB,Maris93} can be studied only
through the non-linear equation. If we could express the
vertex function 
$\tilde \Gamma_\nu(q,p)=S(q) \Gamma_\nu(q,p) S(p)$ as a
functional of the fermion propagator $S$ and solve the SD
equation in terms of $S(p)$, not $S^{-1}(p)$, we could
obtain the solution which includes the non-linear effect
without the linearization and/or the decomposition in the
following sense.
If we start from the SD equation (\ref{SDf}) for
$S$ into which   
\begin{eqnarray}
   S(p) = {\hat{p} X(p) + Y(p) \over p^2},
\end{eqnarray}
is substituted, a pair of integral equations for $X$ and $Y$
is obtained.  In general, although $X$ and
$Y$ couple each other, the equations are still linear in
each variable, $X$ or $Y$.  This is not the case for a pair
of equations for $A$ and $B$ obtained from the
decomposition of the SD equation for $S^{-1}(p)$.  It is
much easier to obtain the solution, $X$ and $Y$, for such
equations which include the non-linear effect through
\begin{eqnarray}
   X(p) := {A(p)p^2 \over A^2(p)p^2+B^2(p)}, 
   \quad
   Y(p) := {B(p)p^2 \over A^2(p)p^2+B^2(p)}.
\end{eqnarray}
Note that, for $B$ small,
\begin{eqnarray}
   X(p) \cong {1 \over A(p)}, 
   \quad
   Y(p) := {B(p) \over A^2(p)}.
\end{eqnarray}
while, for $B$ large
\begin{eqnarray}
   X(p) := {A(p)p^2 \over B^2(p)}, 
   \quad
   Y(p) := {p^2 \over B(p)}.
\end{eqnarray}
The wavefunction renormalization $A(p)$ and the mass
function $M(p)$ are obtained from
\begin{eqnarray}
   M(p) := {B(p) \over A(p)} = {Y(p) \over X(p)}, 
   \quad
   Z(p) := {1 \over A(p)}  
   =  X(p) \left( 1+{M^2(p) \over p^2} \right) .
\end{eqnarray}
\par
In the exact sense, it is impossible to realize such a
situation in general.   However, we find, in 1+1
dimensional case, such a simple situation is realized and
we can solve the SD equation without any approximation.
In this case the SD equation (\ref{SDt}) is linear
in $S$ and can be reduced to a decoupled pair of integral
equations for $X$ and $Y$.  

\subsection{exactly soluble truncated SD equation}
\par
At first glance, the right-hand-side of SD equation
(\ref{SDf}) seems to be quadratic in $S$.  This
observation is not right, because at least the longitudinal
WT identity for the vertex function (\ref{WTL1}) gives a
linear part.  If the transverse part for the vertex is also
linear in $S$, the SD equation might be solved exactly.
In order to examine such a possibility, we try to use the
truncated form of the transverse WT identity up to the same
form as the (\ref{WTT4}).  Substituting the longitudinal WT
identity (\ref{WTL5}) and the truncated transverse WT
identity (\ref{WTT4}) into (\ref{SDf}) by way of
(\ref{integrand}), we get the SD equation for the fermion
propagator (in the chiral limit $M=0$):
\begin{eqnarray}
  \hat{p} S(p) = 1 + ie^2 \int {d^Dk \over (2\pi)^D}
  S(p-k) \hat{k}
  \left[ (D-1) {D_T(k) \over k^2}   - {\xi(k^2) \over k^4}
\right]  ,
\label{SDt}
\end{eqnarray}
where we have used 
$\gamma_\mu \sigma_{\nu\mu} = i(1-D) \gamma_\nu$
and dropped the identically vanishing integral.
This SD equation is exact only in 1+1 space-time dimensions.
The solution is easily found by moving to the coordinate
space, since the Fourier transformation changes the
convolution in momentum space into the simple product in
the coordinate space:
\begin{eqnarray}
  i \hat{\partial} S(x) = \delta^D(x) 
  - ie^2 S(x) \int {d^Dk \over (2\pi)^D}  \hat{k} 
  \left[ {D_T(k) \over k^2}  - {\xi(k^2) \over k^4} \right] 
  e^{-i k \cdot x} .
\end{eqnarray}
Once $D_T$ is given and does not include $S$,
\footnote{
The quenched approximation is the simplest case where
$D_T(k)=1/k^2$.
}
this equation
can be solved exactly and the solution is given by
\begin{eqnarray}
  S(x) = S_0(x)
  \exp \left\{ - i e^2 \int {d^Dk \over (2\pi)^D}
  \left[ {D_T(k) \over k^2} - {\xi(k^2) \over k^4} \right]
  (e^{-i k \cdot x}-1) \right\} ,
  \label{sol}
\end{eqnarray}
where we have assumed the translational invariance
$S(x,y) = S(x-y,0) := S(x-y)$. 
It is obvious that this solution is consistent with the LK
transformation Eq.~(\ref{LKtransf2}).
\par
The above truncation for the vertex is consistent with
the multiplicative renormalizability and the LK
transformation. The detailed
comparison with the decomposition of Ball-Chiu \cite{BC80}
will be given in a subsequent paper
\cite{Kondo96c}, together with the solution of the SD
equation in $D$=2+1 and 3+1 dimensions.
\par

\section{1+1 dimensional case} 
\setcounter{equation}{0}

In $D=1+1$ dimensions, by virtue of the simple transverse
WT identity (\ref{WTT4}) 
and the usual longitudinal WT identity (\ref{WTL5}), we can
write down the exact and closed integral equation for
the SD equation for the fermion propagator $S(x)$ in
the massless bare fermion limit $M=0$,  which can be solved
without approximation.  From the consideration in the
previous section, the exact full fermion propagator of
massless QED$_2$ is given by
\begin{eqnarray}
  S(x) &=& S_0(x)
  \exp \left\{ - i e^2 \int {d^2k \over (2\pi)^2}
  \left[ {D_T(k) \over k^2} - {\xi(k^2) \over k^4} \right]
  (e^{-i k \cdot x}-1) \right\} ,
\nonumber\\
  D_T(k) &=& {1 \over k^2 - {e^2 \over \pi}},
  \label{sol2}
\end{eqnarray}
where the bare fermion propagator is given by
\begin{eqnarray}
S_0(p) := {1 \over \hat{p}}, 
\quad 
S_0(x) = \int {d^2p \over (2\pi)^2} e^{ip \cdot x}S_0(p)
=  {1 \over 4\pi} {\hat{x} \over x^2} .
\end{eqnarray}
In the $\xi=0$ gauge, this result coincides with
Schwinger's result \cite{Schwinger62}. Obviously this
solution obeys the LK transformation. Therefore this
solution gives the exact solution in the arbitrary gauge
$\xi$.  
In fact, it turns out that this solution agrees with the
exact solution obtained in the path integral formalism.
The exact solution of this type is obtained for more
general models in 1+1 dimension, e.g. the gauged Thirring
model \cite{Kondo96a}. 
This issue will be discussed separately
\cite{Kondo96c}.
\par
\par
In massless QED$_2$ (massless Schwinger model), the gauge
field acquires a mass $e/\sqrt{\pi}$ due to radiative
correction which can also be interpreted as a
consequence of anomaly.  It is usually claimed that the
exact solution (\ref{sol2}) has no extra pole at $p \not=
0$ and hence the fermion remains massless.
Nevertheless, we can show that the exact solution does not
exclude the dynamical fermion mass generation in the
chiral limit $M=0$.  This can be shown by solving the
Euclidean version of the SD equation in momentum space
\cite{Kondo96c}.

\section{Conclusion and discussion}
\setcounter{equation}{0}

In this paper, we rederived the transverse WT and the
chiral WT identities based on the path integral formalism
and examined the possible existence of anomaly
for such a new type of WT identities.
In the framework of the SD equation, we have proposed the
strategy in which the transverse WT identity as well as the
usual longitudinal WT identity is used to find the
appropriate ansatz for the vertex function in writing down
the SD equation for the fermion propagator.  
Especially, in $D=2$ dimensions, we have shown that this
strategy can be performed without any approximation and
leads to the exact SD equation for the fermion propagator
which can be exactly solved. 
\par 
It is interesting to try to solve the SD equation for the
fermion propagator in $D > 2$ dimensions under an ansatz
for the vertex function which is guided by the transverse WT
identity (as well as the longitudinal WT identity).   
Such a preliminary attempt has been done already in
\cite{KK88} by truncating the transverse WT identity up to
the nontrivial order in strong coupling QED in 3+1
dimensions.  More detailed investigation based on
the scheme proposed in this paper will be given in a
subsequent work \cite{Kondo96d}.

\newpage

\section*{Acknowledgments}
The author would like to thank Dr. Ian J.R. Aitchison for
kind hospitality in Oxford.
He is also very grateful to Prof. Yashushi Takahashi who
brought his attention to the transverse WT identity in the
very early stage of this work \cite{KK88} in 1988.
This work is supported in part by the Japan Society for the
Promotion of Science and the Grant-in-Aid for Scientific
Research from the Ministry of Education, Science and
Culture (No.07640377).

%\newpage

\end{document}